\documentclass[aps,prl,onecolumn,10pt,noshowpacs,noshowkeys, notitlepage,twoside,a4paper,nobibnotes,superscriptaddress]{revtex4-1}
\usepackage{color}
\usepackage[utf8]{inputenc}
\usepackage{amsmath}
\usepackage{amsfonts}
\usepackage{amssymb}
\usepackage{graphicx}
\usepackage{grffile}

\begin{document}
\title{Energy of a free Brownian particle coupled to thermal vacuum}

\author{J. Spiechowicz}
\affiliation{Institute of Physics, University of Silesia, 41-500 Chorz{\'o}w, Poland}
%
\author{J. {\L}uczka}
\email[Correspondence and requests for materials should be addressed to J.{\L}.]{ (e-mail: jerzy.luczka@us.edu.pl)}
\affiliation{Institute of Physics,  University of Silesia, 41-500 Chorz{\'o}w, Poland}
\email{jerzy.luczka@us.edu.pl} 

\begin{abstract}

Experimentalists have come to temperatures very close to absolute zero at which physics that was once ordinary becomes extraordinary. In such a regime quantum effects and fluctuations start to play a dominant role. In this context we study the simplest open quantum system, namely, a free quantum Brownian particle coupled to thermal vacuum, i.e. thermostat in the limiting case of absolute zero temperature. We analyze the average energy $E=E(c)$ of the particle from a weak to strong interaction strength $c$ between the particle and thermal vacuum. The impact of various dissipation mechanisms is considered. In the weak coupling regime the energy tends to zero as $E(c) \sim c\, \ln{(1/c)}$ while in the strong coupling regime it diverges to infinity as  $E(c) \sim \sqrt{c}$.  We demonstrate it for selected examples of the dissipation mechanisms defined by the memory kernel $\gamma(t)$ of the Generalized Langevin Equation.
We reveal how at a fixed value of $c$ the energy $E(c)$ depends on the dissipation model: one has to compare values of the derivative $\gamma'(t)$ of the dissipation function $\gamma(t)$ at time $t=0$ or at the memory time  $t=\tau_c$ which characterizes the degree of non-Markovianity of the Brownian particle dynamics.
The impact of low temperature is also presented.
\end{abstract}

\maketitle
The journey towards the absolute zero temperature was started in the early 20th century when Heike Kamerlingh Onnes and his colleagues discovered techniques to liquify helium. Nowadays the rapid development of technology made scientists even more eager to reach this temperature in the lab so that racing towards the absolute zero is accelerating swiftly. The lowest temperature currently achieved in laboratories is of the order of picokelvins, i.e. many orders lower than the average temperature of the universe $T = 2.73 \, \mbox{K}$. At these temperatures we gain access to a world of exotic phenomena and physics that was once ordinary becomes extraordinary. Implications of such bizarre properties seemingly are boundless and range from gravitational wave detection, superconductivity, spintronics to quantum computing and other coming technologies.

At low temperature quantum effects start to play a role in which fluctuations are an inherent part. The origin of quantum fluctuations is two-fold: (i) the Heisenberg uncertainty principle and (ii) an environment of temperature $T$ being a source of quantum thermal noise.  However, even at absolute zero temperature $T=0$, there are still vacuum fluctuations that may induce observable effects. Many experiments unveil the role of quantum fluctuations in the ultracold regime. One can mention the motion of macroscopic mechanical objects  \cite{clark}, heat transfer induced by quantum fluctuations between two objects separated by a vacuum gap \cite{chiny}, directly observed reactants, intermediates, and products of bimolecular reactions \cite{hu}, optomechanical systems and mechanical resonators \cite{amir}, glass formation \cite{markland},  quantum control and characterization of charge quantization \cite{pierre}. Another examples of experiments concerning zero-point fluctuations are described e.g. in Refs \cite{silveri,bezginov,lecocq,riek,fragner,tang,casimirE,fin}. These works provide observations of various effects driven by quantum fluctuations in closed and open quantum systems. Apart from the above interest in fundamentals of physics, engineering of the quantum vacuum to create novel devices and protocols for quantum technologies has been developing in recent years \cite{sabin}.

The existence of vacuum fluctuations is one of the most important predictions of modern quantum field theory. One can mention two celebrated examples to evidence it: the Lamb shift \cite{lamb, lamb_physrep} and the Casimir effects \cite{casimir,lif,casimir_revmodphys}. The related phenomenon is the zero-point energy being the lowest possible energy that a quantum mechanical system may have. A well-known example is a quantum harmonic oscillator of frequency $\omega_0$. If it is considered as a closed system then its ground state energy is $(1/2) \hbar \omega_0$. If the oscillator is not perfectly isolated and interacts with thermostat of temperature $T$ then its average energy is $ (1/2) \hbar \omega_0 \coth(\hbar \omega_0/ 2k_BT)$, where $k_B$ is the Boltzmann constant. At absolute zero temperature $T=0$ its energy is $(1/2) \hbar \omega_0$, i.e. the same as for the isolated oscillator. However, it is true only in the limit of weak coupling between the oscillator and thermostat.  If the oscillator-thermostat coupling is not weak then its energy at $T=0$ can be much greater than $(1/2) \hbar \omega_0$. The additional portion of energy comes from thermostat fluctuations.

It is interesting to consider a free quantum particle in this context. Its energy is not quantized and its allowed values are the same as those of a classical counterpart. If it interacts with a heat bath of temperature $T$, then according to the classical statistical mechanics, the average energy is $(1/2)k_BT$ and it tends to zero when $T\to 0$. In the deep quantum regime, its average energy is non-zero even if $T\to 0$. In this paper we revisit this problem. We study the mean energy $E$ of the free quantum particle coupled to thermal vacuum, i.e. thermostat in the limiting regime of  absolute zero temperature $T = 0$. We focus on the impact of interaction strength between the system and thermal vacuum and analyze the role of different dissipation mechanisms. We also discuss fluctuations of energy, the correlation function of thermal vacuum noise  and scaling of the memory kernel of the Generalized Langevin Equation. Finally, we briefly present  the impact of temperature and the harmonic potential.  Appendices contain proofs of asymptotics of the mean energy for strong and weak particle-thermostat coupling for selected examples of the dissipation mechanism.\\

\noindent {\bf Model of a free quantum Brownian particle} \\
We consider the standard model of a free quantum Brownian particle coupled to a heat bath of temperature $T$. For the paper to be self-contained and for the reader's convenience,  we now recall certain basic notions and important elements of this model, see also section Methods and Ref. \cite{bialasPRA}. It is a quantum particle of mass $M$ coupled to a heat bath that is described by the Caldeira-Leggett Hamiltonian, see e.g.  \cite{maga,caldeira,ford,ingol,et2,breuer,ph,weis},
\begin{equation} \label{H}
H=\frac{p^2}{2M} + \sum_i \left[ \frac{p_i^2}{2m_i} + \frac{m_i
\omega_i^2}{2} \left( q_i - \frac{c_i}{m_i \omega_i^2} x\right)^2 \right],   
\end{equation}
where the heat bath is modeled as a set of non-interacting quantum harmonic oscillators.   The operators $\{x, p\}$ are the coordinate and momentum operators of the Brownian particle and $\{q_i, p_i\}$ refer to the coordinate and momentum operators of the $i$-th thermostat oscillator of mass $m_i$ and the eigenfrequency $\omega_i$. The parameter $c_i$ characterizes the coupling between the particle and the $i$-th oscillator. All coordinate and momentum operators obey canonical equal-time commutation relations. 
 
From the Heisenberg equations of motion for all coordinate and momentum operators $\{x, p, q_i,p_i\}$  one can obtain an effective equation of motion for the particle coordinate $x(t)$ and momentum $p(t)$ \cite{bialasEnt}. It is called a generalized quantum Langevin equation and for the momentum operator of the Brownian particle it reads \cite{bialasPRA}
\begin{equation}\label{GLE2}
{\dot p}(t)
+\frac{1}{M} \int_0^t \gamma(t-s) p(s)ds = -\gamma(t) x(0)+ \eta(t),
\end{equation}
where {the dot denotes the derivative with respect to time and $\gamma(t)$ is the memory function (damping or dissipation kernel), 
\begin{equation} \label{gamma}
\gamma(t) =\sum_i \frac{c_i^2}{m_i \omega_i^2} \cos(\omega_i t)  \equiv 
\int_0^{\infty}  J(\omega) \cos(\omega t)  d \omega, \quad   
J(\omega) = \sum_i \frac{ c_i^2}{ m_i \omega_i^2} \delta(\omega -\omega_i) 
\end{equation}
which can be expressed by  the spectral function $J(\omega)$  of the thermostat that contains information on its modes and the Brownian particle-thermostat interaction. {\it  Remark:} The above definition  of the spectral density $J(\omega)$ differs from another frequently used form $\tilde J(\omega) = \omega J(\omega)$.  We prefer the definition as in Eq. (\ref{gamma}) because of a direct relation to the cosine Fourier transform $\hat{\gamma}_F(\omega)$ of the dissipation function (\ref{gamma}),  i.e. $\hat{\gamma}_F(\omega) = J(\omega)$. Here the Ohmic case corresponds to $J(\omega)= const$. The operator $\eta(t)$ can be interpreted as quantum thermal noise acting on the Brownian particle and has the form  
\begin{equation} 
\eta(t) =\sum_i c_i \left[q_i(0) \cos(\omega_i t) + \frac{p_i(0)}{m_i \omega_i}\sin(\omega_i t) \right], 
\label{force} 
\end{equation}
which depends on the thermostat operators $\{q_i(0), p_i(0)\}$ at the initial moment of time.    

One can solve Eq. (\ref{GLE2}) to find $p(t)$ and calculate averaged kinetic energy $E(t)=\langle p^2(t)\rangle/2M$ of the Brownian particle (the notation $\langle \cdot \rangle$ stands here for the averaging over the initial state of the composite system). It is equal to the total average energy of the particle. In the thermodynamic limit of the infinitely extended heat bath and  for $t \to \infty$,  when a thermal equilibrium state is reached, the average kinetic energy $E$ of the Brownian particle can be presented in the form (for a detailed derivation, see Ref. \cite{bialasPRA,superstatistics})
\begin{equation}\label{Ek}
E = \lim_{t\to\infty} \frac{1}{2M} \langle p^2(t)\rangle = \int_0^{\infty} \frac{\hbar \omega}{4} \coth\left({\frac{\hbar \omega}{ 2k_BT}}\right)
\mathbb{P}(\omega)  d\omega
\end{equation}
and 
\begin{eqnarray}\label{P}
\mathbb{P}(\omega) = \frac{1}{\pi} \left[\hat{R}_\mathcal{L}(i\omega)
 + \hat{R}_\mathcal{L}(-i\omega) \right],  
\end{eqnarray}
where 
\begin{equation}\label{RL} 
\hat{R}_\mathcal{L}(z) = \frac{M}{Mz + \hat \gamma_{\mathcal{L}}(z)}, \quad  
\hat \gamma_\mathcal{L}(z) = \int_0^{\infty} e^{-zt} \gamma(t) dt. 
\end{equation}  
The function $\mathbb{P}(\omega)$ fulfils all conditions imposed on the {\it probability density}: (i) it is non-negative, i.e.  $\mathbb P(\omega)\ge 0$, and  (ii) normalized on the positive real half-line, i.e.  $\int_0^{\infty} d\omega \; {\mathbb P}(\omega) = 1$. The corresponding proof is presented in Ref. \cite{bialasJPA}. Eqs. (\ref{Ek})-(\ref{P}) constitute a quantum counterpart of the energy equipartition theorem well known for classical systems. It says that in quantum physics energy is not equally distributed among the degrees of freedom but it is allocated according to the corresponding probability density function $\mathbb{P}(\omega)$. Because the model is exactly solvable the probability density  $\mathbb{P}(\omega)$ obtained from Eq. (\ref{P}) is exact and determined by Eq. (\ref{RL}), i.e. by the Laplace transform $\hat{R}_\mathcal{L}(z)$ of the response function $R(t)$. In turn, Eq. (\ref{RL}) contains the Laplace transform $\hat{\gamma}_{\mathcal{L}}(z)$ of the memory function $\gamma(t)$ in Eq. (\ref{GLE2}) and as such it depends on the spectral function $J(\omega)$, which via Eq. (\ref{gamma}), comprises all information on the oscillator-thermostat interaction and frequencies of the heat bath modes. 

Recently, it has been proven that the relation similar to Eq. (\ref{Ek}) holds true for all  quantum systems for which the concept of kinetic energy has sense (e.g spin systems are outside of this class) \cite{JSP20}. The quantum system can be composed of an arbitrary number of non-interacting or interacting particles, subjected to any confining potentials and coupled to thermostat with arbitrary coupling strength.

In the presently considered case all dynamical quantities are almost periodic functions of time when thermostat consists of a finite number of oscillators. In particular, the dissipation function $\gamma(t)$ is an almost periodic function of time. In order to consistently model the dissipation mechanism, the thermodynamic limit should be imposed by assuming that a number of the thermostat oscillators tends to infinity. Then the dissipation function (\ref{gamma}) decays to zero as $t\to\infty$ and the singular spectral function $J(\omega)$ in Eq. (\ref{gamma}) (which is a distribution rather than an ordinary function) is expected to tend to a (piecewise) continuous function. All what we need to analyze the averaged energy $E$ of the Brownian particle is the memory kernel $\gamma(t)$ in Eq. (\ref{GLE2}) which defines the dissipation mechanism or equivalently the spectral distribution  $J(\omega)$ that contains all information on the particle-thermostat interaction.\\

\noindent {\bf Results: Average  energy of the Brownian particle at zero temperature}\\
At non-zero thermostat temperature $T>0$, the average energy of the free quantum Brownian particle given by Eq. (\ref{Ek}) is always greater than at zero temperature  $T=0$. When $T\to 0$ then $\coth(\hbar \omega/2k_BT) \to 1$ and \mbox{Eq. (\ref{Ek})} reduces to the form 
\begin{equation}\label{Ek0}
 E =  \int_0^{\infty} \frac{\hbar\omega}{4} \, \mathbb{P}(\omega) \, d\omega,  
\end{equation}
which is proportional to the first statistical moment of the probability density 
$\mathbb{P}(\omega)$. It can be interpreted as an averaged kinetic energy 
$\hbar \omega/4$ per one degree of freedom of thermostat oscillators which contribute to $E$ according to the probability distribution $\mathbb{P}(\omega)$. The latter quantity, c.f. Eqs. (\ref{P}) and (\ref{RL}), is defined solely by the dissipation function $\gamma(t)$. The choice of $\gamma(t)$ is arbitrary, although in principle it should be determined by  properties of the environment. As outlined above to guarantee the consistent description $\gamma(t)$ needs to be a bounded and decaying function of time. In the following we consider several examples of $\gamma(t)$ in order to investigate how $E$ depends on $\gamma(t)$ and whether there is an universal behaviour of $E$ which is robust against changes of the dissipation mechanism $\gamma(t)$. 
\begin{figure}[t]
	\centering
	\includegraphics[width=0.5\linewidth]{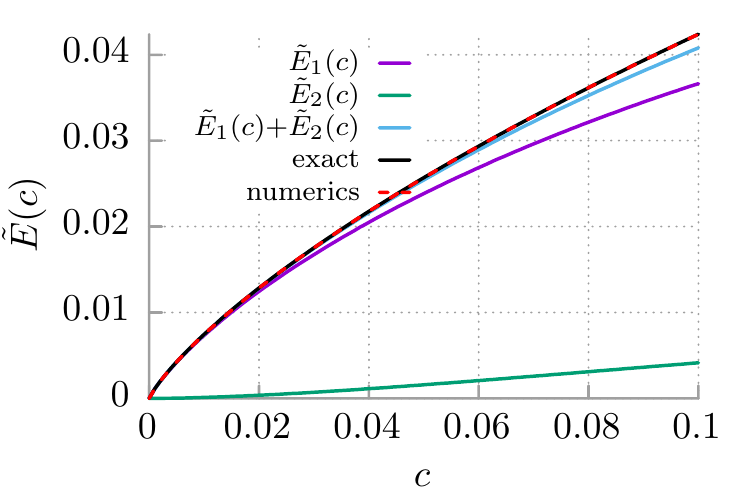}
	\caption{The rescaled average energy $\tilde E(c)$ of the free quantum Brownian particle in the limiting case of weak coupling $c$. The first two leading contributions $\tilde{E}_1(c)$ and $\tilde{E}_2(c)$ to $\tilde E(c)$ are depicted.}
	\label{fig1}
\end{figure} 

\noindent {\bf Analytically tractable case: Drude model. } 
The so-called Drude model is defined by the exponentially decaying damping function or/and the spectral density given by the following form \cite{weis} 
\begin{equation}
	\label{g-drude}
	\gamma_D(t) = \gamma_0 \, e^{-t/\tau_c}, \quad J_D(\omega) = \frac{2}{\pi}\frac{\gamma_0\tau_c}{1 + \tau_c^2\omega^2}, 
\end{equation}
where $\gamma_0 >0$ is the particle-thermostat coupling strength and $\tau_c > 0$ is the memory time which characterizes the degree of non-Markovianity of the Brownian particle dynamics. Its inverse $\omega_c =1/\tau_c$ is the Drude cutoff frequency.  
The probability distribution is found to be \cite{bialasPRA}
\begin{eqnarray} \label{P_D}
	\mathbb{P}_D(\omega) = \frac{2}{\pi} \frac{M \gamma_0}{\tau_c} \, \frac{1}{(M\omega^2-\gamma_0)^2 +(M\omega/\tau_c)^2}
\end{eqnarray}
and the mean energy of the Brownian particle is given by the formula
\begin{eqnarray}\label{EkD}
 E =  \frac{1}{2\pi} \frac{M\gamma_0}{\tau_c} \int_0^{\infty} \frac{\hbar \omega}{(M\omega^2-\gamma_0)^2 +(M\omega/\tau_c)^2} \, d\omega.  \nonumber\\
 \ \ \ \ 
\end{eqnarray}
We note that there are three parameters of the system $\{M, \gamma_0, \tau_c\}$. 
The dimensionless quantities can be introduced as follows 
\begin{equation} \label{times}
\tilde E =  \frac{\tau_c E}{\hbar}, \quad
x = \tau_c \omega,  \quad   c= \frac{\gamma_0 \tau_c^2}{M},   
\end{equation}
which transform the relation (\ref{EkD}) to the form 
\begin{eqnarray}\label{zero}
\tilde E = \tilde E(c) = \frac{1}{2\pi} \int_0^{\infty} \frac{c x}{(x^2 - c)^2 + x^2} \,dx.  
\end{eqnarray}
In this scaling the parameter $c$ is the dimensionless particle-thermostat coupling strength. It is impressive  that now the system is completely characterized by only one parameter $c$. The above integral (\ref{zero}) can be explicitly calculated yielding quite remarkable expression for the mean energy, namely,  
\begin{eqnarray} 
\tilde E(c) &=& \frac{c}{4\pi \sqrt{1-4c}} \ln{ \frac{1-2c+\sqrt{1-4c}}{1-2c-\sqrt{1-4c}}} , \; c < 1/4,  \label{weak} \\
&=& \frac{c}{2\pi \sqrt{4c-1}} \left[ \frac{\pi}{2} +  \arctan{\frac{2c-1}{\sqrt{4c-1}}} \right], \; c > 1/4,  \label{strong} \\
&=& \frac{1}{4\pi}, \quad c= \frac{1}{4} \label{mean}.  
\end{eqnarray}
The dependence of $\tilde E(c)$ upon the coupling constant $c$ is depicted in Fig. 1. It is a monotonically increasing function of the latter parameter. If  $c\to 0$ then $\tilde E(c) \to 0$ and $\tilde E(c) \to \infty$ when  $c \to\infty $.  
In the weak coupling regime $c \ll 1$, the first two leading contributions to the energy have the form
\begin{eqnarray}\label{weak0}
 \tilde{E}(c) =   \tilde{E}_1(c) +  \tilde{E}_2(c), \quad
 \tilde{E}_1(c) =  \frac{c}{2\pi} \ln(1/c), \quad 
 \tilde{E}_2(c)=  \frac{c^2}{\pi}[\ln(1/c) -1].   
\end{eqnarray} 
Their graphical representation is also depicted in \mbox{Fig. 1}. The term $\tilde{E}_1(c)$ is already known in the literature \cite{weis}. It is worth noting that the leading order contribution to the Lamb shift is also logarithmic and reads $\alpha^5 \ln(1/\alpha)$, where $\alpha$ is a fine-structure constant. The correction $\tilde{E}_2(c)$ is the next to the leading order contribution to $\tilde E(c)$ for small $c$. The term $(-c^2/\pi)$ is included to minimize the deviation from the exact value of the zero-point particle energy. 
We now return to the dimensional variables and the leading order contribution to the dimensional energy is 
\begin{equation}\label{dimE}
 E_1 = \hbar \omega_c\, \tilde{E}_1 =  \frac{\hbar}{2\pi} \, \frac{\gamma_0 \tau_c}{M} 
 \ln\left(\frac{M}{\gamma_0 \tau_c^2}\right).  
\end{equation} 
It is the purely quantum term which is proportional to $\hbar$ and tends to zero when the coupling constant $\gamma_0 \to 0$ or the memory time $\tau_c \to 0$ or the particle mass $M\to \infty$. The asymptotics of $\tilde E(c)$ can be evaluated also for the limit of strong coupling. By inspecting (\ref{strong}) we find that
\begin{eqnarray}\label{asymp}
 \tilde E(c)  \sim \sqrt{c}, \quad c \gg 1,  
\end{eqnarray}
i.e. it increases with the coupling constant as a square root of $c$.  
In Appendix A we prove that the same asymptotics holds true for non-zero temperatures, $T>0$. \\

\begin{figure*}[t]
	\centering
	\includegraphics[width=0.32\linewidth]{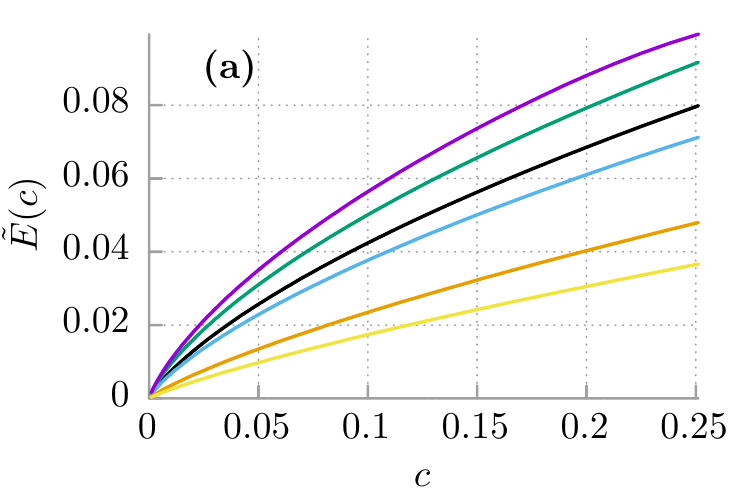}
	\includegraphics[width=0.32\linewidth]{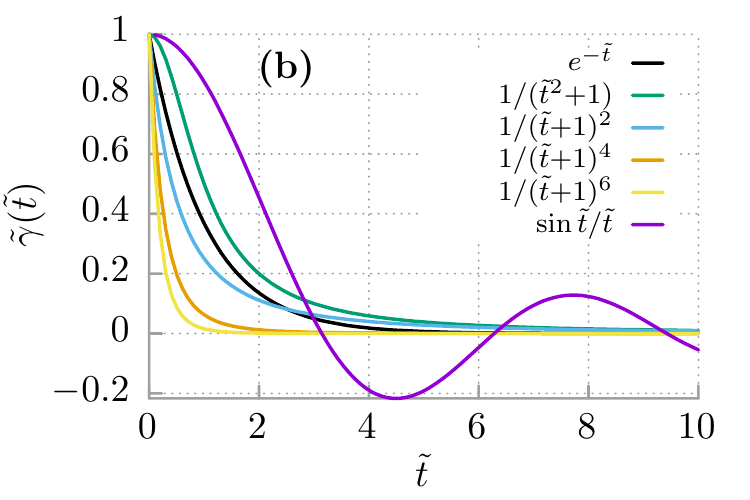}
	\includegraphics[width=0.32\linewidth]{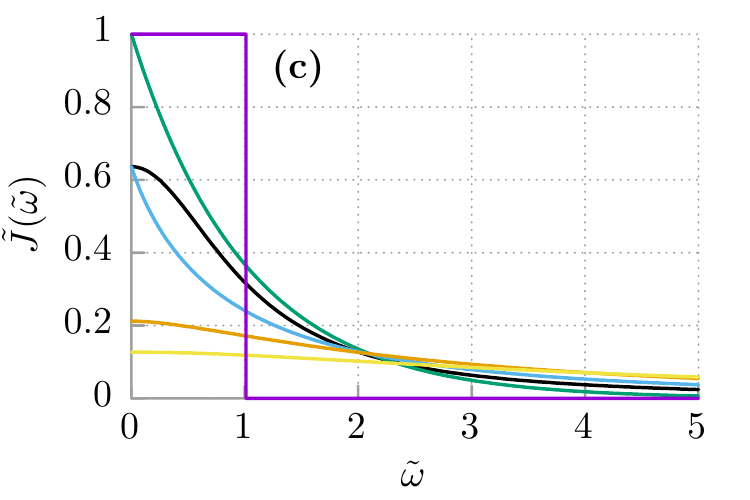}\\
	\includegraphics[width=0.32\linewidth]{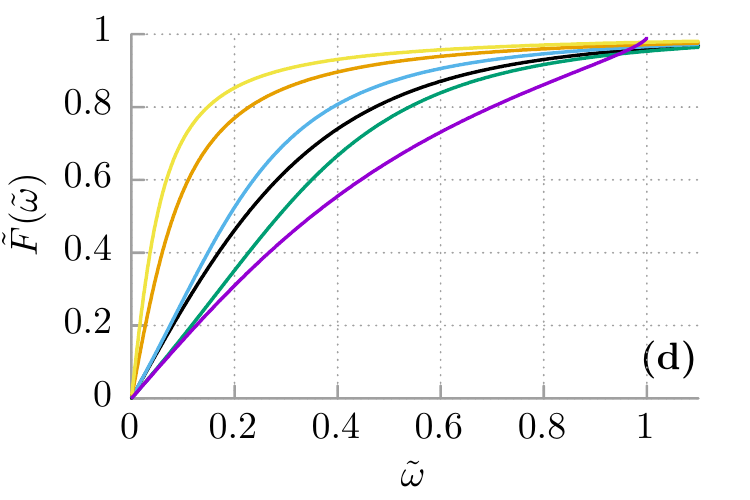}
	\includegraphics[width=0.32\linewidth]{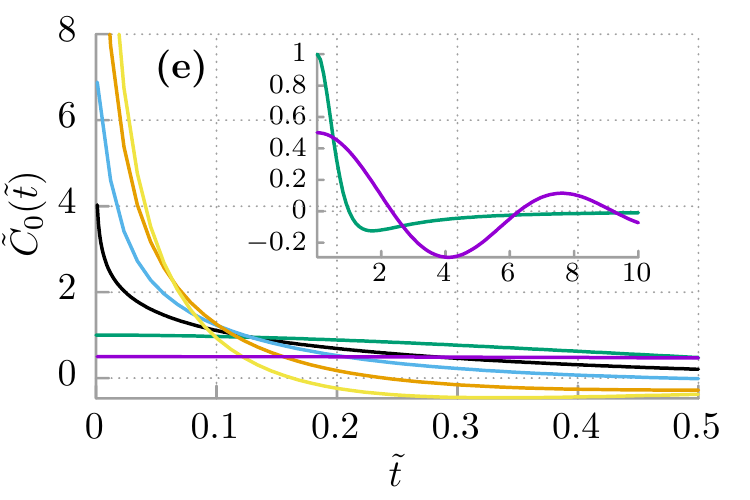}
	\caption{(a): The average energy of the quantum Brownian particle $\tilde E(c)$ depicted for different dissipation mechanisms.
		(b): the  dissipation kernel $\tilde{\gamma}(\tilde{t})$, 
		(c): the spectral density $\tilde{J}(\tilde{\omega})$, 
		(d): the cumulative distribution function 	$\tilde{F}(\tilde{\omega})$ 
	and (e): the correlation function $\tilde{C}_0(\tilde{t})$ of quantum noise, in inset we present the magnified correlation functions for the Debye and Lorentz case.
	The dimensionless quantities are: $\tilde{\gamma}(\tilde{t}) = \gamma(\tilde{t}/\omega_c)/\gamma_0$, $\tilde{J}(\tilde{\omega})= (\omega_c/\gamma_0) J(\omega_c \tilde{\omega})$, $\tilde{F}(\tilde{\omega}) = F(\omega)$ and 
	$\tilde{C}_0(\tilde{t}) = 2 C_0(\tilde{t}/\omega_c)/(\gamma_0 \hbar \omega_c)$.  
	The dimensionless variables are: $\tilde{t}=\omega_c t$ and $\tilde{\omega}=\omega/\omega_c$. In panel (d) $c = 0.25$.}
	\label{fig2}
\end{figure*}
\noindent {\bf Other  examples of the dissipation mechanism. } 
We now want to analyze how  the average energy of the quantum Brownian particle 
depends on different dissipation mechanisms modeled by $\gamma(t)$ and check the interrelations between the corresponding zero-point energies.

 {\it 1. Lorentzian decay.  } 
 As the second example we pick the Lorentz type dissipation for which 
\begin{equation}
	\label{g-cauchy}
	\gamma_L(t) = \gamma_0\, \frac{1}{1+ (t/\tau_c)^2}, \quad  
	J_L(\omega) = \gamma_0 \tau_c \,e^{-\tau_c \omega}.
\end{equation}
Such a choice of the dissipation kernel leads  to the following probability
distribution  
\begin{equation}
	\label{p-cauchy0}
	\mathbb{P}_L(\omega) = 
	\frac{4 \nu_0 \, e^{- \tau_c \omega}}{\pi^2 \nu_0^2 \, e^{-2\tau_c \omega} + h^2(\omega)}, \quad \nu_0 = \frac{\gamma_0 \tau_c}{M}, 
\end{equation}
where 
\begin{eqnarray}
	h(\omega) = 2 \omega + \nu_0 e^{\tau_c \omega} \mbox{Ei}(-\tau_c \omega) - \nu_0 e^{-\tau_c \omega} \mbox{Ei}(\tau_c \omega), \quad  
	\mbox{Ei}(z) = \int_{-\infty}^z \frac{e^{t}}{t} \, dt
\end{eqnarray}
and $\mbox{Ei}(z)$ is the exponential  integral. 
For this mechanism of dissipation the  mean energy $E$ in Eq. (\ref{Ek0}) cannot  be calculated analytically. However, in Appendix B, we evaluate the strong coupling asymptotics and demonstrate that it is the same as for the Drude model, i.e. 
$E \sim \sqrt{c}$ for $c \gg 1$. 

{\it 2. Family of algebraic decay. } 
 This class of dissipation mechanisms is defined by the following formula for the memory kernel and the spectral density, 
\begin{equation}
	\label{g-algebraic-n}
	\gamma_{n}(t) = \gamma_0 \frac{1}{(1+ t/\tau_c)^n}, \quad 
	J_{n}(\omega) = \frac{\gamma_0 \tau_c}{\pi} \left[e^{-i\tau_c \omega} \mbox{E}_n(-i\tau_c \omega) + e^{i\tau_c \omega} \mbox{E}_n(i \tau_c \omega)\right], 
	\quad \mbox{E}_n(z) = \int_1^\infty dt\, \frac{e^{-zt}}{t^n}, 
\end{equation}
where $n \in \mathbb{N}$,  $n \geq 2$ and 
$\mbox{E}_n(z)$ is the exponential integral. 
The probability distribution takes the form 
\begin{eqnarray}    
\label{p-algebraic-n}
\mathbb{P}_{n}(\omega) = \frac{\nu_0}{\pi} \,
\frac{ e^{-i\tau_c \omega} \mbox{E}_n(- i \tau_c \omega) + e^{i \tau_c \omega}\mbox{E}_n(i \tau_c \omega)}{\left[\omega + i \nu_0 e^{-i \tau_c \omega} \mbox{E}_n(-i \tau_c \omega)\right]\left[\omega - i\nu_0 e^{i \tau_c \omega} \mbox{E}_n(i\tau_c \omega)\right]}. 
\end{eqnarray}

{\it 3. The Debye-type model. }  
 Another example of the dissipation model reads
\begin{equation}
	\label{g-sin}
	\gamma_S(t) = \gamma_0 \,	\frac{\sin{(t/\tau_c)}}{t/\tau_c}, \quad 
	 J_S(\omega) = \frac{\gamma_0}{\omega_c} \, \theta (\omega_c - \omega), 
\end{equation}
where $\omega_c=1/\tau_c$ is the cut-off frequency and $\theta(x)$ denotes the Heaviside step function.  This model of dissipation is peculiar: the spectral density  
is a positive constant on the {\it compact} support $[0,\omega_c]$ determined by the memory  time $\tau_c$ and is zero outside this interval of frequencies. 
 Under this assumption the probability density can be presented as  
%
\begin{eqnarray}
	\label{p-sin}
		\mathbb{P}_S(\omega) = \frac{4M}{\gamma_0} \,
	\frac{\omega_c \,  \theta(\omega_c-\omega)}{ \pi^2 + \left[\ln (\omega_c+\omega)  - \ln (\omega_c-\omega)  - 2M\omega_c \,\omega/\gamma_0 \right]^2} 
	\end{eqnarray}
%
and has  the same  compact support $[0, \omega_c]$ as the spectral function $J_S(\omega)$. The corresponding integral (\ref{Ek0}) for the mean energy $E$ cannot be analytically calculated with the probability distribution (\ref{p-sin}).  However, in Appendix C, we evaluate the weak coupling regime and show that it is the same as for the Drude model, namely, $E \sim c \,\ln(1/c)$ for $c \ll 1$. \\

\noindent {\bf Average  energy vs  dissipation mechanism. }
In Fig. 2 (a) we present dependence of the average  energy $\tilde{E}(c)$ on the particle-thermostat coupling strength $c$ for different forms of the dissipation mechanism. To facilitate the analysis,  we plot the damping kernel $\gamma(t)$ and  the spectral density $J(\omega)$  in panels (b) and (c), respectively. In panel (e) we display  the correlation function $C_0(t)$ of quantum noise (\ref{force}) [see Eqs. (\ref{correl}) and (\ref{correl0})]. The reader can immediately note that the sequence (from the top to the bottom) of the zero-point energy curves $\tilde E(c)$ for different dissipation mechanisms is the same as the ordering  of the damping kernels $\gamma(t)$ and the spectral densities $J(\omega)$ for small times $t$ and frequencies $\omega$, respectively. 
In contrast, it is rather difficult to reveal any universal pattern in the impact of the dissipation form on the corresponding correlation function $C_0(t)$ of quantum thermal noise $\eta(t)$, see panel (e) of Fig. 2. Similarly, there is no evident relation between the probability densities $\mathbb{P}_j(\omega), (j= D, S, L, n=2, 4, 6)$ (not depicted) and  the zero-point energy curve $\tilde E(c)$. 
However, it is instructive to analyze the cumulative distribution function $F_j(\omega)$, namely, 
\begin{eqnarray} \label{cumul} 
F_j(\omega) = \int_0^{\omega} \mathbb{P}_j(u) du, \quad j= D, S, L, n=2, 4, 6.
\end{eqnarray} 
It is depicted  in Fig. 2 (d) from which it follows that the correlation between  $F_j(\omega)$ and  $\tilde E(c)$ is evident: 
If the cumulative distribution function is greater then the zero-point energy $\tilde E(c)$  is smaller.  If for two probabilities $F_j(\omega) > F_l(\omega)$ for $\omega \in (0, \omega_c/2)$ then for the corresponding energies $\tilde E_j(c) < \tilde E_l(c)$. 
The above observations allow us to formulate the following conjectures:
\begin{enumerate} 
\item The decay rate  of the damping kernel $\gamma(t)$ crucially modify the energy $\tilde E$. If $\gamma(t)$ decreases rapidly then $\tilde E$ is small. In other words, if  $\gamma_1(t) < \gamma_2(t)$ for $t\in (0, \tau_c)$ then  $E_{\gamma_1} < E_{\gamma_2}$. 
\item If the main contribution to the zero-point energy $\tilde E$ comes from  the environment  oscillators of small frequencies $\omega$ then $\tilde E$ is small. 
It means that if 
$J_1(\omega) < J_2(\omega)$ for
$\omega \in (0, \omega_c)$  then $E_{J_1} < E_{J_2}$.
\item There is no non-zero lower bound for the zero-point energy $\tilde E(c)$ of the free quantum Brownian particle,  i.e. for any $\gamma_i(t)$ one can find  $\gamma_j(t)$ that $E_j < E_i$. 
\end{enumerate}
By analyzing Fig. 2 we find three quantifiers which allow to order the sequence of the  energy curves for various dissipation mechanisms. They are: the memory kernel $\gamma(t)$ or the spectral function $J(\omega)$, or the cumulative distribution function $F_j(\omega)$. Perhaps the most convenient way to arrange them is by inspecting the derivative $\gamma'(t)$ of the memory kernel $\gamma(t)$ at zero $t=0$ or at the memory time  $t=\tau_c$. These values are listed in Table \ref{tabela}. The rule is the following: If $\gamma'(0)$  decreases then the mean energy $E$ also decreases. In turn, if $\gamma'(\tau_c)$ increases then $E$ decreases. The only exception is the case of the Debye dissipation function which, however, belongs to a different class than the rest of the considered models. Indeed, the Debye spectral density $J_S(\omega)$ possesses a compact support $[0, \omega_c]$ while the remaining spectral densities are non-zero on  the frequency interval $[0, \infty)$. \\

\renewcommand{\arraystretch}{2}
\begin{table}[t]
\centering
\begin{tabular}{|c|c|c|}

\hline 
• & $\tilde{\gamma}'(0)$ & $\tilde{\gamma}'(1)$ \\ 
\hline 
Debye $\tilde{\gamma}_S(\tilde{t}) = \sin{\tilde{t}}/\tilde{t}$ & 0 & -0.301169 \\ 
\hline 
Lorentz $\tilde{\gamma}_L(\tilde{t}) = 1/(\tilde{t}^2 + 1)$ & 0 & -0.5 \\ 
\hline 
Drude $\tilde{\gamma}_D(\tilde{t}) = \exp{(-\tilde{t})}$ & -1 & -0.367879 \\ 
\hline 
Algebraic $n=2$, $\tilde{\gamma}_2(t) = 1/(\tilde{t} + 1)^2$ & -2 & -0.25 \\ 
\hline 
Algebraic $n=4$, $\tilde{\gamma}_4(t) = 1/(\tilde{t} + 1)^4$& -4 & -0.125 \\ 
\hline 
Algebraic $n=6$, $\tilde{\gamma}_6(t) = 1/(\tilde{t} + 1)^6$ & -6 & -0.046875 \\ 
\hline
\end{tabular}
\caption{Numerical values of the derivatives $\tilde{\gamma}'(\tilde{t})$ of various 
dimensionless dissipation functions computed for $\tilde{t} = 0$ and $\tilde{t} = 1$, i.e. for the memory  time $t = \tau_c$ which characterizes  the degree of non-Markovianity of the particle dynamics.}
\label{tabela}
\end{table} 

\noindent {\bf Discussion} \\
\noindent {\bf Fluctuations of energy. } 
In order  to analyze fluctuations of energy let us note that 
 in the stationary state the Brownian particle momentum depends linearly on thermal noise  $\eta(t)$ (cf. Eq. (\ref{p(t)}) in the section Methods),  
\begin{equation}
\lim_{t \to \infty} p(t) = \lim_{t \to \infty} \int_0^t R(t-u)\eta(u)du.
\end{equation}
Statistical characteristics of quantum thermal noise  $\eta(t)$ are analogous to a classical stationary Gaussian stochastic process.  For the above  reasons the particle momentum $p$ is also Gaussian implying that
\begin{equation}
\langle p^4 \rangle = 3\langle p^2 \rangle^2. 
\end{equation}
From this relation it follows that fluctuations of energy are proportional  to the average energy $E$.  Indeed, the energy variance is  $(\Delta E)^2 = 2 E^2$ and in consequence the  standard deviation of energy  is proportional to the average energy, $\Delta E = \sqrt{2}  E$. Therefore the dependence of energy fluctuations $\Delta E$ on the coupling constant $c$ is exactly the same as for $E$.  In particular, $\Delta E$ tends to zero for $c\to 0$ and it diverges when $c\to \infty$. \\

\noindent {\bf The correlation function of thermal vacuum  noise. }
For classical systems the correlation function \mbox{$C(t)=C_{cl}(t)$} of thermal noise $\eta(t)$ is equal, up to a constant factor, to the damping function $\gamma(t)$. Indeed, for high temperature 
\begin{equation} \label{highTcorrel2}
 \coth \left(\frac{\hbar \omega}{2k_B T}\right) \approx \frac{2k_B T}{\hbar\omega} 
\end{equation}
and from Eq. (\ref{correl}) it follows that 
\begin{equation} \label{clas}
C_{cl}(t) = k_B T \gamma(t). 
\end{equation}
Properties of $C_{cl}(t)$ can be deduced from  Fig. 2 (b). At absolute zero temperature $T=0$ its quantum counterpart $C_0(t)$ is obtained from Eq. (\ref{correl})  and reads 
\begin{equation} \label{correl0}
C_0(t) = \int_0^{\infty} \frac{\hbar \omega}{2} J(\omega)\cos(\omega t) d\omega.
\end{equation}
In contrast, it is not proportional to $\gamma(t)$ as in the classical case.  Representative examples of $C_0(t)$ are depicted in Fig. 2 (e). For the Drude model, the correlation function (\ref{correl0})   reads 
\begin{eqnarray}
	\label{correl-drude}
	C_D(t) = -\frac{\gamma_0}{\pi} \frac{\hbar \omega_c}{2} \, \left[\mbox{e}^{-\omega_c t}\,  \mbox{Ei}(\omega_c t) + \mbox{e}^{\omega_c t}\, \mbox{Ei}(-\omega_c t)\right]. 
\end{eqnarray}
 When $t\to 0$ then $C_D(t) \to \infty$ and the second moment of  noise diverges,  $\langle \eta^2(t) \rangle \to \infty$.
For the Debye-type model, it is bounded and has the form  
\begin{eqnarray} \label{correlS}
C_S(t) = \gamma_0  \frac{\hbar \omega_c}{2} \, \left[\frac{\sin(\omega_c t)}{\omega_c t} + 
\frac{\cos(\omega_c t) -1}{(\omega_c t)^2}\right],  \quad  
\langle \eta^2(t) \rangle = C_S(0)= \gamma_0 \frac{\hbar \omega_c}{4}
\end{eqnarray}
and for the Lorentzian decay it is also bounded,  
\begin{equation} \label{correlL}
C_L(t) = \gamma_0 \frac{\hbar \omega_c}{2} \, \frac{1-(\omega_c t)^2}{[1+(\omega_c t)^2]^2}, \quad
\langle \eta^2(t) \rangle = C_L(0) = \gamma_0 \frac{\hbar \omega_c}{2}.  
\end{equation}
For the algebraic decay of $\gamma(t)$ given by Eq. (\ref{g-algebraic-n}) there is no an analytical expression for $C_0(t)$. Its numerical calculation is presented in Fig. 2 (e). For all members of the family of algebraic decay the second moment of noise does not exist, $\langle \eta^2(t) \rangle = \infty$. 

There are three crucial disparities: (i) In the classical case $C_{cl}(t) \to 0$ for $T\to 0$. In the quantum case $C_0(t) \ne 0$ at absolute zero temperature $T=0$. (ii) $C_0(0)$ can diverge for quantum systems while its classical counterpart $C_{cl}(0)$ has to be finite, cf. Eq. (21). (iii) if $C_{cl}(t)$ is positive then $C_{0}(t)$ may assume negative values.  It means that quantum  noise can exhibit negative correlations (anticorrelations) while its classical counterpart exhibits only positive ones. For tailored forms of the dissipation kernels classical noise may also be anticorrelated as it is the case e.g. for the Debye model. \\

\noindent {\bf Scaling of the memory kernel. }
In this paper, we choose the memory kernels in such a way that all have the same value at the initial time, $\gamma(0)=\gamma_0$. 
In the literature,  the memory kernel $\gamma(t)$ is frequently defined in such a way that it tends to the Dirac delta distribution $\delta(t)$ when the memory time $\tau_c$ tends to zero, i.e. as a Dirac $\delta$-sequence (cf. Ref \cite{bialasPRA}). E.g. for the Drude model the most common form reads  
\begin{equation}
	\label{drude2}
	\gamma(t) = \frac{\gamma}{\tau_c} \, e^{-t/\tau_c}.
\end{equation}
Indeed,  $\lim_{\tau_c \to 0} \gamma(t) = \gamma \delta(t)$ and for the integral part of the Langevin equation (\ref{GLE2}) one gets  
\begin{equation}\label{integ}
 \frac{1}{M} \int_0^t \gamma(t-s) p(s) \, ds  \to \gamma \,  \frac{p(t)}{M}.   
\end{equation}
In this limit, the integro-differential equation (\ref{GLE2}) reduces to the differential Langevin equation. It is often called the white noise limit or Markovian approximation. Let us verify its consequences. Firstly, according to Eq. (\ref{gamma2}), in such a case the force constant $k_0 = \gamma(0) = \gamma/\tau_c$.
%
When $\tau_c \to 0$ then $k_0 \to \infty$ and the counter-term in Eq. (\ref{counter}) becomes greater and greater. 
Secondly, the zero-point energy of the  Brownian particle tends to infinity. It is explicitly seen from Eq. (\ref{dimE}) by inserting $\gamma_0 = \gamma/\tau_c$. Indeed, $E_1 = \hbar \omega_c \tilde{E}_1  \propto  \ln(1/\tau_c) \to \infty$. Moreover, if $\tau_c$ is varied as a control parameter then the force constant $k_0=\gamma/\tau_c$ is modified and the Hamiltonian (\ref{H}) is altered. In this way  one compares e.g. the average energy $E$ for two different values of $\tau_c$, i.e. for two different Hamiltonians (namely for two different physical systems).  It shows that the problem of the white noise limit or the Markovian approximation in quantum physics is subtle and still not satisfactory resolved. \\

\begin{figure*}[t]
	\centering
	\includegraphics[width=0.45\linewidth]{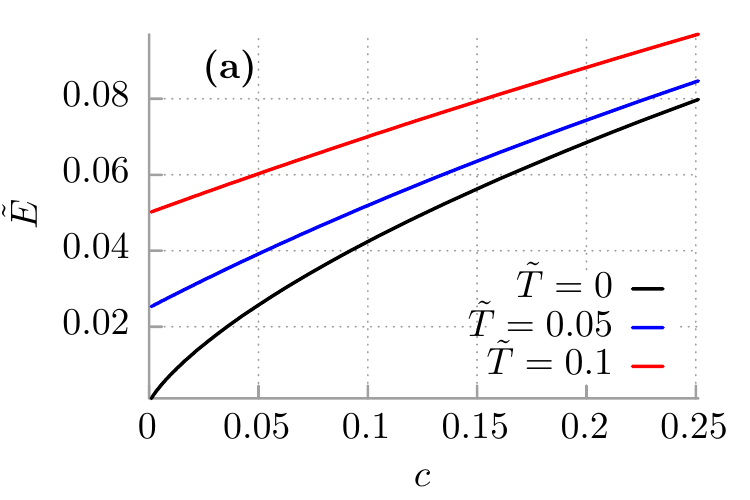}
	\includegraphics[width=0.45\linewidth]{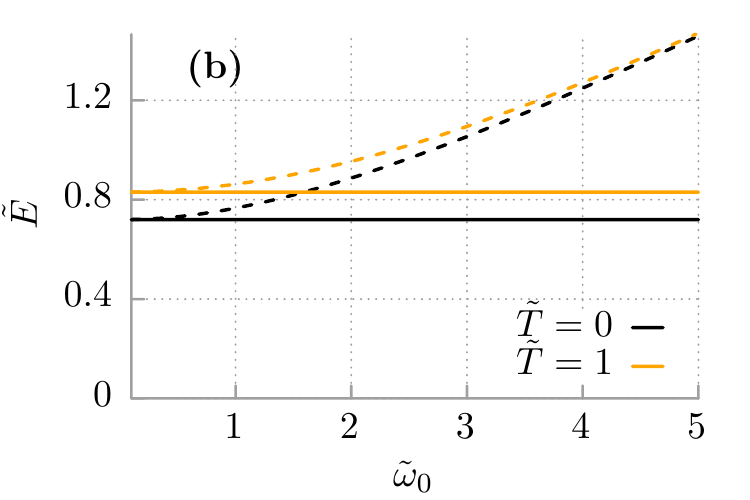}
	\caption{Panel (a): Impact of temperature on the dimensionless mean kinetic energy $\tilde{E}$  of the free quantum particle within the Drude model. Panel (b): Influence of the eigenfrequency $\tilde{\omega}_0$ of the harmonic oscillator on its mean  kinetic energy $\tilde{E}$. Solid lines represent the results for the free particle where the dashed ones correspond to the harmonic oscillator. The dimensionless energy $\tilde{E}  = E/\hbar \omega_c$, temperature $\tilde{T} = k_B T/\hbar \omega_c$ and $\tilde{\omega} = \omega/\omega_c$. In panel (b) $c = 10$.}
	\label{fig3}
\end{figure*}

\noindent {\bf Impact of temperature and potential energy. }
In order to complement the analysis, in Fig. 3 we show the influence of temperature and a potential on the average {\it kinetic energy} of the quantum Brownian particle. As an example we present the case of a harmonic oscillator for which the potential is  $U(x) = M\omega_0 x^2/2$. It is an exactly solvable model \cite{bialas}. As expected, if temperature of a thermal bath increases the average kinetic energy of the particle grows as well. It is obvious that the average potential energy becomes greater when the eigenfrequency $\omega_0$ increases. The same hold true for the total energy. What is in clear contrast to classical result is the dependence of the average kinetic energy on the eigenfrequency $\omega_0$. Here, the kinetic energy  grows together with $\omega_0$ whereas classically it is independent of the latter parameter and equal to $(1/2)k_B T$ as the equipartition theorem states. Solid lines represent the results for the free particle where the dashed ones correspond to the harmonic oscillator. \\

\noindent {\bf Conclusions} \\
We have revisited the paradigmatic model of a free quantum Brownian particle in contact with quantum thermostat in the limiting case of  absolute zero temperature  and studied the mean  energy of the particle. We have scrutinized the impact of a limited class of  dissipation mechanisms for which behaviour of the zero-point energy of the Brownian particle as a function of the rescaled coupling strength between the system and the thermostat is similar. 

We show that the sequence  of the average  energy curves $E(c)$ for different dissipation mechanisms is  the same as the sequence of the damping curves $\gamma(t)$,   the spectral densities $J(\omega)$ and the cumulative distribution functions $F(\omega)$  for small values of their arguments, respectively.  In particular, we find out that the best quantifier is the derivative $\gamma'(t)$ of the dissipation function $\gamma(t)$ at time $t=0$ or at  the characteristic time  $t=\tau_c$. 
For the Drude model we additionally obtained an exact analytical formula for the zero-point energy of the free Brownian particle. It allowed us to evaluate the asymptotic forms of  the energy in the limit of weak and strong particle-environment coupling at zero and non-zero temperature. The Debye model exhibits the same weak coupling asymptotics as the Drude model. From Fig. 2(a) it follows that  also  for the Lorentzian decay the same weak coupling asymptotics holds true.  Moreover, the Lorentz model displays the same strong coupling asymptotics as the Drude model.  

We briefly discussed the problem of energy fluctuations $\Delta E$. However, because they are proportional to the average energy $E$, their functional behavior is the same as $E$.   In particular, $\Delta E$ tends to zero for $c\to 0$ and it diverges when $c\to \infty$. 
We compared the correlation functions of thermal noise in the classical and quantum case. In particular, quantum thermal noise can exhibit negative correlations (anticorrelations) while its classical counterpart exhibits only positive ones. 
We pointed out some subtleties and imperfections of the discussed model when the damping kernel is scaled in such a way that it tends to the Dirac delta distribution. When the memory time approaches zero, the force constant as well as the zero point energy tend to infinity. 
Last but not least, we discussed the influence of the harmonic potential on the zero-point energy of the particle. Finally, we have to emphasize that the presented results and statements are correct for a broad but limited class of examples of the memory function (or the spectral density). Still there is an open question how general the results are.   \\

\noindent {\bf Methods} \\
In order to calculate the average kinetic energy $E$ given by Eq. (\ref{Ek}) one has to solve Eq. (\ref{GLE2}) to find $p(t)$. Because  Eq. (\ref{GLE2}) is a linear integro-differential equation it  can be solved by e.g. the Laplace method. The result reads
\begin{equation}\label{p(t)} 
p(t) = R(t) p(0) - x(0)\int_0^t R(t-u)\gamma(u) du  
+ \int_0^t R(t-u) \eta(u) du,  
\end{equation}
where $R(t)$ is called a response function and is determined by its Laplace transform 
$\hat{R}_\mathcal{L}(z)$, see Eq. (\ref{RL}). Having $p(t)$ one can calculate the symmetrized momentum-momentum correlation function which, in the thermodynamic limit imposed on a heat bath, is expressed by the symmetrized noise-noise correlation function \cite{bialasPRA}.
The statistics of noise $\eta(t)$ defined in  Eq. (\ref{force}) is crucial  for evaluation of $E$. 
We assume the factorized initial state  of the composite  system, i.e., $\rho(0)=\rho_S\otimes\rho_B$, where $\rho_S$ is an arbitrary state of the Brownian particle and $\rho_B$ is the canonical Gibbs state of the heat bath  
of temperature $T$, namely,
\begin{eqnarray} \label{gibbs}
\rho_B = \mbox{exp}(-H_B/k_B T)/\mbox{Tr}[\mbox{exp}(-H_B/k_B T)],
\quad H_B= \sum_i \left[ \frac{p_i^2}{2m_i} + \frac{1}{2} m_i \omega_i^2 q_i^2 \right],  
\end{eqnarray}
where $H_B$ is the Hamiltonian of the heat bath. The factorization means that there are no initial correlations between the particle and  thermostat. The initial preparation turns the force $\eta(t)$ into the operator-valued quantum thermal  noise which in fact is a family of non-commuting operators whose commutators are $c$-numbers. This noise is unbiased and its mean value is zero, 
$\langle \eta(t) \rangle = \mbox{Tr} \left[ \eta(t) \rho_B\right] = 0.$
Its symmetrized correlation function 
\begin{equation} \label{correl}
C(t-s)=   \frac{1}{2} \langle   \eta(t)\eta(s)+\eta(s)\eta(t)\rangle
=
 \int_0^{\infty} \frac{\hbar \omega}{2} \coth \left(\frac{\hbar \omega}{2k_B T}\right) J(\omega)\cos[\omega(t-s)] d\omega
\end{equation}
depends on the time difference. 
The higher order correlation functions are expressed  by $C(t_i-t_j)$  and have the same form as statistical characteristics for classical stationary Gaussian stochastic processes. Therefore $\eta(t)$  defines  a quantum stationary Gaussian process with time homogeneous correlations. 

The next quantity which we should consider is 
the counter-term in the Hamiltonian (\ref{H}), i.e. the term proportional to $x^2$ (for the relevant discussion, see e.g. Ref. \cite{weis}), 
\begin{equation} \label{counter}
\frac{1}{2} \sum_i  \frac{c_i^2}{m_i \omega_i^2}\, x^2	= \frac{1}{2} k_0 x^2, \quad  k_0 = \sum_i  \frac{c_i^2}{m_i \omega_i^2} = \int_0^{\infty}  J(\omega) d \omega  < \infty.
\end{equation}
The force constant $k_0$ is related to the dissipation function by the relation (\ref{gamma}) from which it follows that 
\begin{equation} \label{gamma2}
\gamma(t) = \int_0^{\infty}  J(\omega) \cos(\omega t)  d \omega, \quad \gamma(0)= k_0 < \infty.  
\end{equation}
It is quite natural that quantities like the force constant $k_0$ and the mean energy $E$ should be finite.  We note that $k_0$ is related to the dissipation function $\gamma(t)$ at time $t=0$ and therefore $\gamma(t)$ as a decaying function of time should be finite, $\gamma(t) < \infty$.  Moreover, from (\ref{counter}) it follows that the spectral density $J(\omega)$  has to be integrable on the positive half-line and the integral is associated with the dissipation function $\gamma(t)$ at the initial moment of time $t=0$. Frequently it is assumed that under some limiting procedure the memory kernel $\gamma(t)$ tends to the Dirac delta in order to study a Markovian regime.  It means that $\gamma(t)$ is an integrable function on the half-axis $t \ge 0$.  We also assume this restriction.
The question is whether the noise correlation function $C(t)$ in Eq. (\ref{correl}) should be finite for all values of time, in particular $C(0)$ which is related to the second moment $\langle \eta^2(t) \rangle$ of thermal noise. It is well known that in classical statistical physics thermal noise is frequently represented as Gaussian white noise for which the second moment does not exist and it is not a drawback. 
One can  keep this question open as long as it does not lead to divergences of relevant measurable observables. \\

\noindent{\bf Appendix A. Strong coupling for the Drude dissipation at $T>0$}\\
For the Drude model, the average energy (\ref{Ek}) of the Brownian particle coupled to thermostat of  non-zero temperature  has the form 
\begin{eqnarray}\label{zero9}
\tilde E = \tilde E(c) = \frac{1}{2\pi} \int_0^{\infty} \frac{c x \coth(Ax)}{(x^2 - c)^2 + x^2} \,dx,  
\end{eqnarray}
where the dimensionless quantities are defined in Eq. (\ref{times}) and 
$A=\hbar \omega_c /2 k_B T$ with  $\omega_c =1/\tau_c$. It corresponds to Eq. (\ref{zero}) for $T>0$. 
We want to evaluate the asymptotics of  (\ref{zero9}) for $c \to \infty$.   From the graph of $x \coth(Ax)$ it follows that for any  number $b>0$ the function $x \coth(Ax) \le b$ for $x\le b/2$. We put $b=2 c^{1/3}$. Next, we note that for $c \gg 1$  the following inequalities hold true: 
\begin{equation}\label{non1}
 x \coth(Ax) \le 2 c^{1/3}  \quad \mbox{for} \quad  x <  c^{1/3},  
\end{equation}
\begin{equation}\label{non18}
 (x^2-c)^2 + x^2 \ge (c- x^2)^2 \ge \left(\frac{c}{2}\right)^2 \quad  \mbox{for} \quad  x \le c^{1/3}. 
\end{equation}
We present the integral in Eq. (\ref{zero9}) as a sum of two integrals,  
\begin{eqnarray}\label{eek}
 \tilde E(c) = \frac{1}{2\pi} \, \int_0^{c^{1/3}} 
 \frac{c x \coth(Ax)}{(x^2 - c)^2 + x^2} \,dx    + 
 \frac{1}{2\pi} \, \int_{c^{1/3}}^{\infty} 
 \frac{c x \coth(Ax)}{(x^2 - c)^2 + x^2} \,dx. 
\end{eqnarray}
If $c \to \infty$ the first integral tends to zero: 
\begin{eqnarray}\label{I0}
  \int_0^{c^{1/3}} 
 \frac{c x \coth(Ax)}{(x^2 - c)^2 + x^2} \,dx  \le
    \int_0^{c^{1/3}} 2 c^{1/3} \, c \, \left(\frac{c}{2}\right)^{-2} dx 
    =  8 \, c^{-1/3}  \to 0. 
 \end{eqnarray}
Now, we consider the second integral. We note that for sufficiently large $c \gg 1$,  
\begin{equation}\label{nierow}
 1 \le \coth(Ax) \le \coth(A c^{1/3}) \quad \mbox{for} \quad x\in (c^{1/3}, \infty) \ \ 
 \end{equation}
and hence  
\begin{eqnarray}\label{I1}
\int_{c^{1/3}}^{\infty}
 \frac{c x}{(x^2 - c)^2 + x^2} \,dx    \le 
 \int_{c^{1/3}}^{\infty}  
 \frac{c x \coth(Ax)}{(x^2 - c)^2 + x^2} \,dx  \le 
\int_{c^{1/3}}^{\infty} 
 \frac{c x \coth(A c^{1/3})}{(x^2 - c)^2 + x^2} \,dx.  \nonumber\\
\end{eqnarray}
 The first integral from the left side (the lower bound) can be analytically evaluated (cf. Eq. (\ref{zero})) and   behaves as $\sqrt{c}$ when  $c \to \infty$. The integral in the second line (the upper bound) also behaves as $\sqrt{c}$  when $c\to \infty$. From the squeeze theorem it follows that the middle integral also  behaves  as 
 $\sqrt{c}$. We conclude that in the case of the strong particle-thermostat coupling 
\begin{eqnarray}\label{asymp2}
 \tilde E(c)  \sim \sqrt{c}, \quad c \gg 1
\end{eqnarray}
holds true both for zero and non-zero temperature in the Drude model of dissipation. \\

\noindent {\bf Appendix B. The Lorentzian decay: strong coupling } \\
We perform the analysis of the strong coupling limit $(c \gg 1)$ in two steps. In the first step, we 
consider the probability density (\ref{p-cauchy0}) in the form 
\begin{equation}
	\label{cauchyx}
	P_c(x) =\frac{1}{\tau_c} \mathbb{P}_L\left(\frac{x}{\tau_c}\right) =  
	\frac{4  e^{-x}}{c[\pi^2  \, e^{-2x} + g^2(x)]}, 
\end{equation}
where 
\begin{equation} \label{gx}
	g(x) = 2x/c - f(x), \qquad f(x) = e^{-x} \mbox{Ei}(x) -  e^{x} \mbox{Ei}(-x).
\end{equation}
Note that $f(x)$ does not depend on the parameter $c$. 
\begin{figure}[t]
	\centering
	\includegraphics[width=0.45\linewidth]{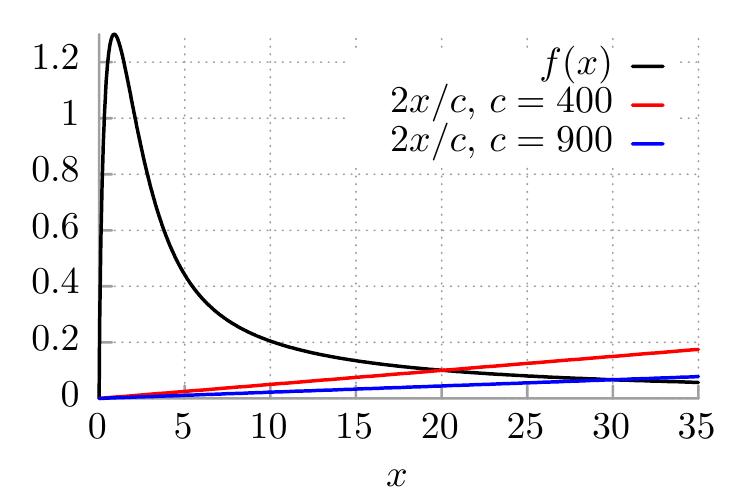}
	\includegraphics[width=0.49\linewidth]{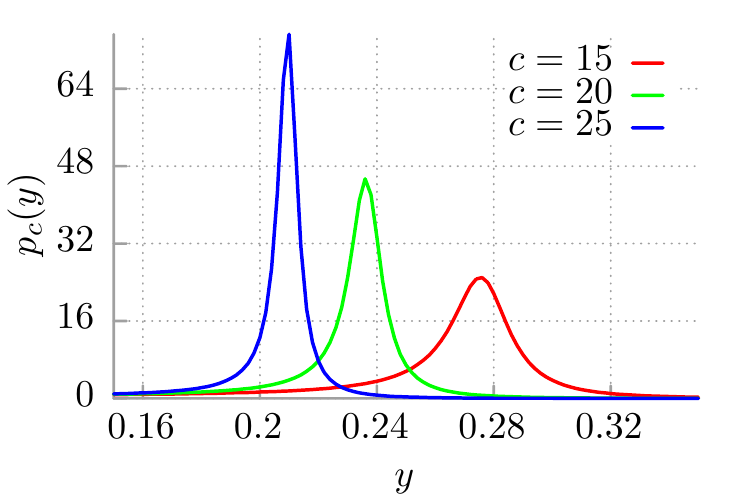}
	\caption{Left panel. Black: the function $f(x)$ defined in Eq. (\ref{gx}).  Red and blue: the straight line $2x/c$ for $c=400$ and $c=900$, respectively. The intersection of the  curves is a root of the equation $g(x)=0$ defined in Eq. (\ref{gx}). Right panel. Example of three terms of the Dirac $\delta$-sequence  
		$p_c(y)$ defined in Eq. (\ref{p-L}).  }
	\label{fig4}
\end{figure}
We analyze $P_c(x)$ on two intervals $x \in (0, \infty) = (0, X_0]\cup (X_0, \infty)$ for some number $X_0$ which depends on $c$ and  is sufficiently smaller than the non-zero root of the function $g(x)$, i.e. $X_0 \ll x_m$, where $x_m$ is a root of the equation $g(x_m)=0$.  On the interval $(0, X_0]$ the density $P_c(x)$ tends to zero and the average energy tends to zero when $c\to \infty$. On the interval 
$(X_0, \infty)$, the density $P_c(x)$ tends to the Dirac delta distribution when $c \to \infty$. Now, we provide analytical arguments indicating how
to isolate the Dirac delta contribution. In Fig. \ref{fig4}, we depict the graph of $f(x)$. 
For any $c>0$ the function $g(x)$ always has a non-zero root $x=x_m$, i.e. $g(x_m)=0$, see Fig. \ref{fig4}. If $c$ increases, the value $x_m$ also increases. For very large $c$, the value $x_m$ is large and the denominator in Eq. (\ref{cauchyx}) is small. In consequence, the density (\ref{cauchyx}) has a peak  
at $x=x_m$ and reads
\begin{equation}
	P_c(x_m) = \frac{4}{c\pi^2} e^{x_m}.  
\end{equation}
Because for large $c$ the value of $x_m$ is also large, we can evaluate how $x_m$ depends on $c$. To this aim we use the asymptotic expansion \cite{grad}  
\begin{equation}
	f(x_m)  \approx \frac{2}{x_m} + \frac{4}{x_m^3} = \frac{2x_m}{c}. 
\end{equation}
Hence 
\begin{equation}
	x_m^2 = \frac{c}{2} [1+ \sqrt{1+8/c} \; ] \approx c \quad  \mbox{for} \quad c \gg 1
\end{equation}

We observe that $x_m$ grows with $c$ as $x_m \sim \sqrt{c}$ and  at this value the probability density is 
\begin{equation}
	P_c(x_m) = \frac{4}{c\pi^2} e^{\sqrt{c}} \to \infty  \quad  \mbox{for} \quad c \to \infty
\end{equation}
This is the part of $P_c(x)$ which  tends to the Dirac delta distribution.  
In the second step, we use a different scaling and  present  the dimensionless energy in the form 
\begin{equation}
	\label{E-L}
	\tilde E(c) = 
	\frac{1}{4} \int_0^{\infty} c y \, p_c(y) \; dy, 
\end{equation}
where the normalized probability density $p_c(y)$ takes the form
\begin{equation}
	\label{p-L}
	p_c(y)=  \frac{4}{\pi} \; \frac{\pi  \, e^{- cy} } 
	{(\pi e^{-cy})^2 + b^2(cy)}, \qquad  
	b(cy) = 2y + e^{cy} \,\mbox{Ei}(-cy) -  e^{-cy} \,\mbox{Ei}(cy). 
	\end{equation}
	It resembles the Dirac delta sequence (see also \cite{imparato}): 
$\epsilon/[\epsilon^2 + x^2]  \to 
\pi \delta(x) \quad \mbox{when} \quad \epsilon \to 0$.
In the strong coupling regime, when $c \gg 1$, the probability density (\ref{p-L}) tends to the Dirac $\delta$-distribution, namely, 
\begin{equation}
	\label{p20}
	p_c(y)  \to  4\, \delta(b(cy)) =  4\, \frac{\delta(y-y_0)}{|b'(y_0|}. 
	\end{equation}
In Fig. \ref{fig4} we visualize three terms of this Dirac $\delta$-sequence.  The value $y_0$ 
is obtained from the equation $b(y_0) = 0$ and for large $c$ it  takes the form $b(y) = 2(y-1/cy)$. Hence $y_0=1/\sqrt{c}$ and $b'(y_0)=4$. Inserting (\ref{p20}) into (\ref{E-L}) yields the asymptotics 
\begin{equation}
	\label{as-L}
	\tilde E(c) \sim \sqrt{c}, \qquad c \gg 1, 
\end{equation}
which is the same as for the Drude model.   

\noindent {\bf Appendix C. The Debye-type model: weak coupling} \\
For the Debye memory function (\ref{g-sin}) the dimensionless zero-point energy reads 
\begin{eqnarray}\label{zero1}
 \tilde E(c) =  \int_0^{1} \frac{c x \, dx}{c^2 \pi^2 +\{2x - 
 c \ln[(1-x)/(1+x)]\}^2 }. \ \ \ \ 
\end{eqnarray}
The dimensionless quantities are defined in Eq. (\ref{times}). In the limit of weak coupling, $c\ll1$,  it can be well approximated by the equation 
\begin{eqnarray}\label{zero2}
 \tilde E(c) \approx   \int_0^{1} \frac{c x \, dx}{\pi^2 c^2 + 4x^2 } 
 =  \frac{c}{8} \ln\left[ 1 + \left(\frac{2}{\pi c}\right)^2\right] \sim c \ln(1/c). 
\end{eqnarray}
It has the same asymptotics as for the Drude model of dissipation. \\


\ \ \\
\noindent {\bf Acknowledgement}\\
The work was supported by the Grant NCN 2017/26/D/ST2/00543. We thank Ryszard Rudnicki (Institute of Mathematics, Polish Academy of Sciences) for his suggestion on the proof presented  in Appendix A. \\
\noindent {\bf Author Contributions}\\
 Both authors contributed extensively to the planning, interpretation, discussion and writing up of this work. \\

\noindent {\bf Competing financial interests} \\
The authors declare no competing financial interests.

\end{document}